\documentclass[]{spie}  

\usepackage{amsmath,amsfonts,amssymb}
\usepackage{graphicx}
\usepackage[colorlinks=true, allcolors=blue]{hyperref}
\usepackage{multirow,bigdelim}

\def\CN2{\mbox{$C_N^2 \ $}}

\def\CT2{\mbox{$C_T^2 \ $}}

\def\sigmal2{\mbox{$\sigma ^{2}_{I} \ $}}

\title{Towards the forecast of atmospheric parameters and optical turbulence above an astronomical site on 24h time scale
} 

\author[a,b]{Gianluca Martelloni}
\author[a]{Elena Masciadri}
\author[a]{Alessio Turchi}
\affil[a]{INAF - Osservatorio Astrofisico di Arcetri, L.go E. Fermi 5, 50125  Florence, Italy}
\affil[b]{INSTM - National Interuniversity Consortium of Materials Science and Technology, Via della Lastruccia 3-13, 50019 Sesto Fiorentino (Florence), Italy}

\authorinfo{Correspondence to Gianluca Martelloni: martelloni@arcetri.astro.it}

\begin{document} 
\maketitle 

\begin{abstract}
Forecast of the atmospheric parameters and optical turbulence applied to the ground-based astronomy is very crucial mainly for the queue scheduling. So far, most efforts have been addressed by our group in developing algorithms for the optical turbulence ($\CN2$) and annexed integrated astroclimatic parameters and quantifying the performances of the Astro-Meso-Nh package in reconstructing such parameters. Besides, intensive analyses on the Meso-Nh performances= in reconstructing atmospheric parameters relevant for the ground-based astronomy has been carried out. Our studies referred always to the night time regime. To extend the applications of our studies to the day time regime, we present, in this contribution, preliminary results obtained by comparing model outputs and measurements of classical atmospheric parameter relevant for the ground-based astronomy in night and day time. We chose as a test case, the Roque de los Muchachos Observatory (Canary Islands), that offers a very extended set of measurements provided by different sensors belonging to different telescopes on the same summit/Observatory. The convective regime close to the ground typical of the day time is pretty different from the stable regime characterising the night time. This study aims therefore to enlarge the domain of validity of the Astro-Meso-Nh code to new turbulence regimes and it permits to cover the total 24 hours of a day. Such an approach will permit not only an application to solar telescopes (e.g. EST) but also applications to a much extended set of scientific fields, not only in astronomical context such as satellite communications. 
\end{abstract}



\section{INTRODUCTION}
\label{sec:intro} 
In this preliminary work we study the performances of the Meso-NH (Mesoscale Non-Hydrostatic model) in analyzing the classical atmospheric parameters relevant for astronomical applications at the surface layer on the Canary observatory sites (Roque de Los Muchachos). Here, our study is limited to wind speed, temperature and relative humidity. 
Firstly we opportunely configure the model orography to carry out the best performance of the model. In particular we test the three and four-domain (3D and 4D) configuration, corresponding to the four resolution of 10km, 2.5km, 0.5km and 0.1km. We perform the simulation by means of initialization data obtained by general circulation model of ECMWF. Hence we focus our attention on the comparison between model outputs and measurements taken by the observatory instrumentations with the aim of validating the numerical predictions of atmospheric parameters. As mentioned above we remark that we study the performance of the model both on night and day time.\\
In this preliminary study we use MESO-NH (Lafore et. al. 1998 [\cite{lafore98}], Lac et al. 2018 [\cite{lac2018}]) model developed by the Laboratoire d'Aerologie, CNRM and Meteo France. The optical turbulence will be treated with the Astro-MESO-NH  module (Masciadri et. al. 1999 [\cite{masciadri99a}]) developed to provide forecast of optical turbulence parameters.\\
The Astro-MESO-NH model is in constant evolution since 1999. It was used in previous studies performed on many telescope installations, such as San Pedro Martir - Mexico (Masciadri et. al. 2004 [\cite{masciadri2004}]), Antarctica (Lascaux et. al. 2009-2010 [\cite{lascaux2009,lascaux2010}]), Mount Graham in Arizona, site of the Large Binocular Telescope (Hagelin et. al. 2010-2011 [\cite{hagelin2010,hagelin2011}]). More recently, the model forecasts for atmospheric and astroclimatic parameters were part of a large validation campaign conducted within the MOSE project, commissioned by the European Southern Observatory (ESO) in order to prove the feasibility of an automated forecast system for their installations in Cerro Paranal and Armazones (VLT and E-ELT respectively) (Masciadri et. al. 2013 [\cite{masciadri2013}], Lascaux et. al. 2013-2015[\cite{lascaux2013,lascaux2015}]). In that context, a new algorithm for the optical turbulence has been proposed and the Astro-Meso-Nh code has been validated on an extended number of nights (several tens) Masciadri et al. 2017 [\cite{masciadri2017}].\\
In this paper we will show the preliminary results of an ongoing validation study on 24h time scale. The study is performed on a limited sample of nights and days, which was used to test the performance of the model and tune the settings in order to obtain the most efficient configuration for a future operational tool.\\

\section{MEASUREMENTS AND TEST SITE CONFIGURATION}
\label{sec:obs} 
\begin{figure}[ht]
\begin{center}
\includegraphics[height=5cm]{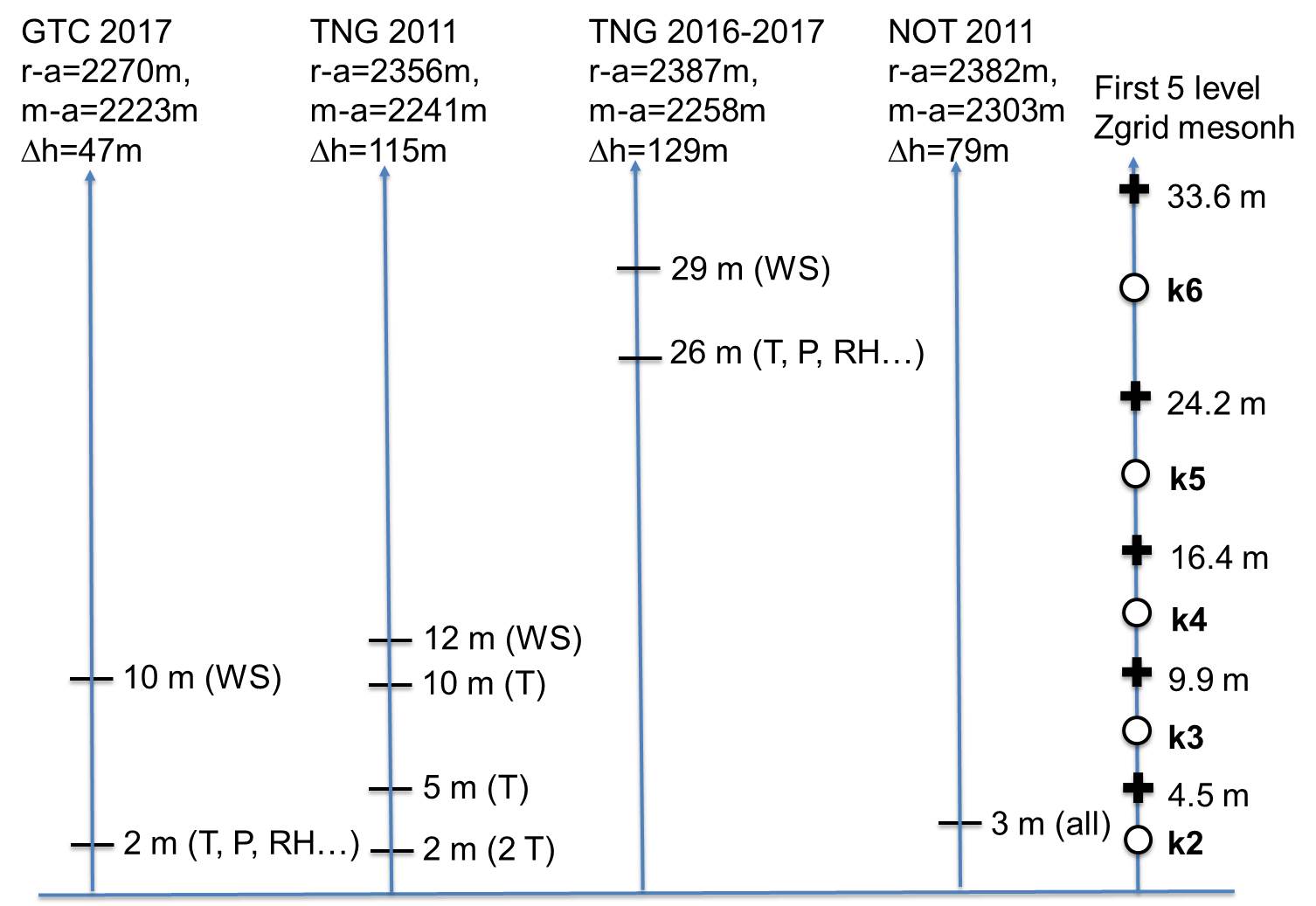}
\end{center}
\caption[example]{\label{fig:wethstat}: Weather stations and sensors used for the comparison between observations and model simulations. We use the data from GTC, i.e., Moradas weather station for year 2017, from NOT for year 2011 and from TNG for year 2011 (Vaisala station) and for years 2016-2017 (Davis station). In the figure the height of sensors and the corresponding levels of  Meso-NH model are indicated. On the right we have the vertical grid of the Meso-Nh model. The circles indicates the centre of the unit cells and the crosses indicate the top border of the unit cell. } 
\end{figure}

Measurements of atmospheric parameters close to the ground have been taken with sensors of weather stations of different telescopes located at the Roque de Los Muchachos Obsvertory. In some cases, the sensors are located on the dome of the telescope (TNG). This large number of references allows us to verify the performances of Meso-NH on a large dataset. The three telescopes of which we use the correspondent atmospheric measures for our study are the GTC (Gran Telescopio Canarias $[28.756611 N, -17.891889 W]$), the TNG (Telescopio Nazionale Galileo $[28.754000 N, -17.889056 W]$) and NOT (Nordic Optical Telescope $[28.757278 N, -17.885083 W]$). The heights of the sensors above the ground is not necessarily the same for all the weather stations and we had no simultaneous measurements for all the three telescopes. We therefore sampled two different periods (2011 and 2017) in order to have at least two different measurements to compare to the models outputs. In Fig.\ref{fig:wethstat} we report a draw in which are shown the heights at which the sensors of the weather stations are located for the comparison with model simulations. We also indicate the real altitude of each telescope and the altitude provided by the orography generated by the model. All stored data are sampled with a time step of one minute, except the data provided from TNG for year 2011 (Vaisala station) that are sampled with a time step of 30 seconds. For our analysis we apply to raw data a moving average of one hour and we re-sample data at 20 minutes (see validation section). In Table \ref{tab:dates2011} and in \ref{tab:dates2017} we report the dates of selected 'nights' for the preliminary validation. The selected 'days' are referred to correspondent previous dates).

\begin{table}[ht]
\caption{Selected nights on 2011 (UT) used for preliminary validation. The selected days on 2011 are associated to the previous dates.} 
\label{tab:dates2011}
\begin{center}       
\begin{tabular}{|c|c|c|c|c|c|c|} 
\hline
\rule[-1ex]{0pt}{3.5ex}  2011/01/15 & 2011/01/17 & 2011/02/14 & 2011/02/28 & 2011/03/03 & 2011/06/26 & 2011/06/30\\
\hline
\rule[-1ex]{0pt}{3.5ex}  2011/07/12 & 2011/08/01 & 2011/08/08 & 2011/08/31 & 2011/09/11 & 2011/09/29 & 2011/10/03\\
\hline
\rule[-1ex]{0pt}{3.5ex}  2011/10/09 & 2011/10/25 & 2011/11/08 & 2011/11/18 & 2011/12/04 & 2011/12/15 & 2011/12/19\\
\hline
\rule[-1ex]{0pt}{3.5ex}  2011/12/31 & & & & & &  \\
\hline
\end{tabular}
\end{center}
\end{table}

\begin{table}[ht]
\caption{Selected nights 2017 (UT) used for preliminary validation. The selected days on 2017 are associated to the previous dates.} 
\label{tab:dates2017}
\begin{center}       
\begin{tabular}{|c|c|c|c|c|c|c|} 
\hline
\rule[-1ex]{0pt}{3.5ex}  2017/01/03 & 2017/01/25 & 2017/02/28 & 2017/03/02 & 2017/03/24 & 2017/04/23 & 2017/04/30\\
\hline
\rule[-1ex]{0pt}{3.5ex}  2017/05/04 & 2017/05/21 & 2017/06/03 & 2017/06/24 & 2017/07/06 & 2017/07/22 & 2017/08/08\\
\hline
\rule[-1ex]{0pt}{3.5ex}  2017/08/30 & 2017/09/13 & 2017/09/28 & 2017/10/03 & 2017/10/19 & & \\
\hline
\end{tabular}
\end{center}
\end{table}

\section{MODEL CONFIGURATION}
\label{sec:mod_conf} 

In order to compute atmospheric simulations we use a mesoscale model called Meso-Nh\footnote{\url{http://mesonh.aero.obs-mip.fr/mesonh/}} [\cite{lac2018}], developed by the CNRM and Laboratoire d'Aerologie (Tolouse, FR) with which we simulate the time evolution of the atmospheric parameters in a volume defined over a finite geographic area around the point of interest. The vertical mesh is defined by a grid over the Gal-Chen and Sommerville coordinate systems [\cite{chen}]. The physics contained in the model is defined on the anelastic version of hydrodynamic equations, which is fundamental to filter acoustic waves. We also consider a one-dimensional mixing length as defined in Bougeault et. al. 1989 [\cite{Bougeault89}] with a one-dimensional 1.5 closure scheme [\cite{Cuxart00}]. The Interaction Soil Biosphere Atmosphere scheme [\cite{Noilhan89}] is used to model the exchange between surface and atmosphere.\\
We used data from the General Circulation Model of ECMWF (European Center for Medium Weather Forecasts), which is extend on the whole globe, to initialize the Meso-Nh model. The GCM has an horizontal resolution of about 9~km (Since March 2016). In previous studies on nighttime operating telescope installations [\cite{turchi2017, lascaux2015}] we simulated only the nighttime portion of each date. In this contribution we treat also the day-time regime. In order to use specifically optimized set-ups for different physical regimes (with quite different convection strength) we decided to run separately two simulations, one for the daytime and one for the nighttime. The daytime simulations is initialized on each date at 6:00 UT and cover 15 hours until 21:00UT. The nighttime simulation is initialized at 18:00UT and also runs for 15 hours up to 9:00 UT. In either case a forcing with new data is performed every 6 hours.\\
To obtain an optimal temporal and spatial resolution we use a grid-nesting configuration [\cite{Stein00}]. Such a technique make use of multiple imbricated domains, each subsequently defined on a smaller total surface with a higher horizontal resolution, while the vertical resolution is the same for all domains. We report in Table \ref{tab:resol} the domain chosen for this study. Each domain is centered on the GTC telescope.\\

\begin{figure} [ht]
\begin{center}
\begin{tabular}{cc} 
\includegraphics[height=5cm]{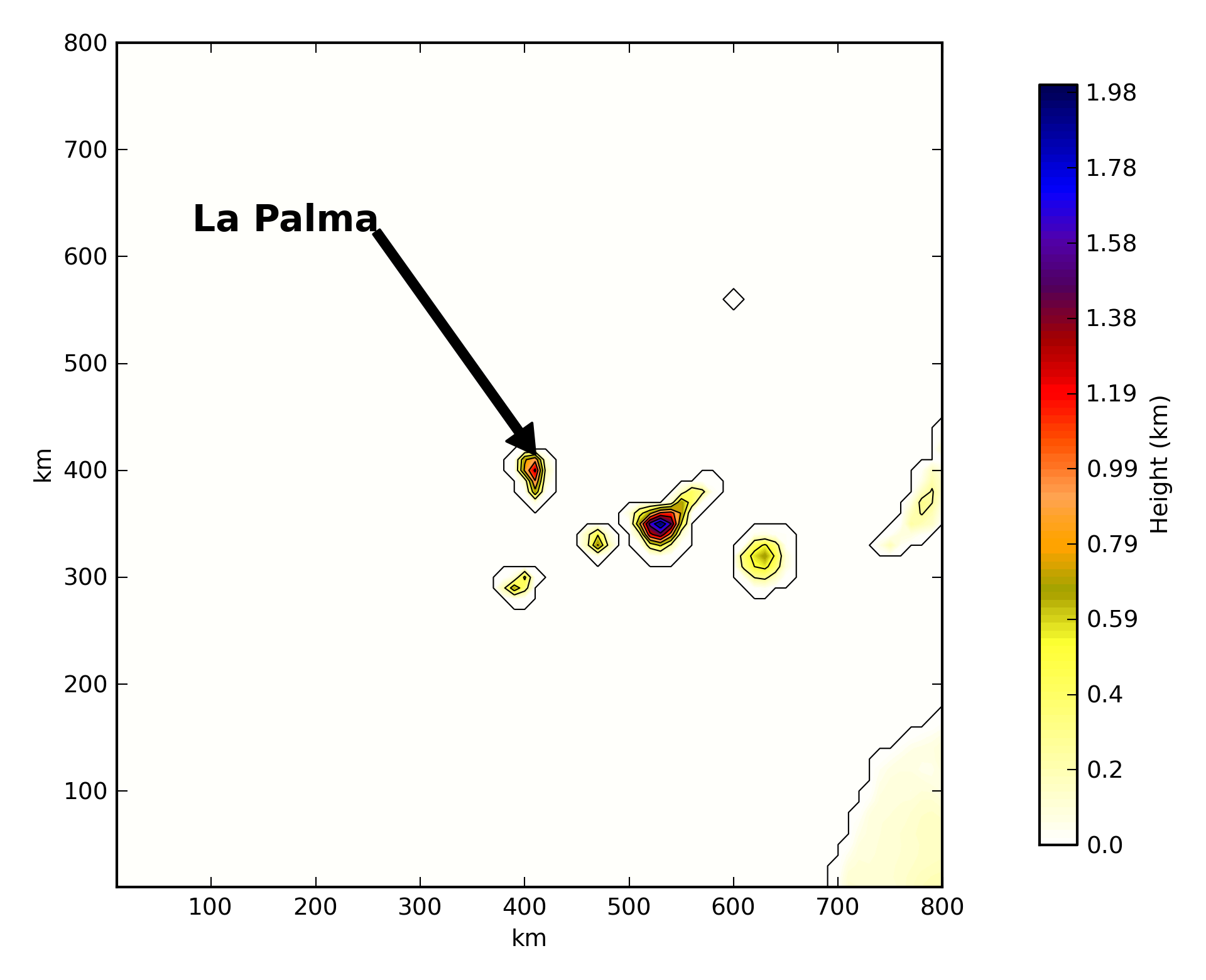} & \includegraphics[height=5cm]{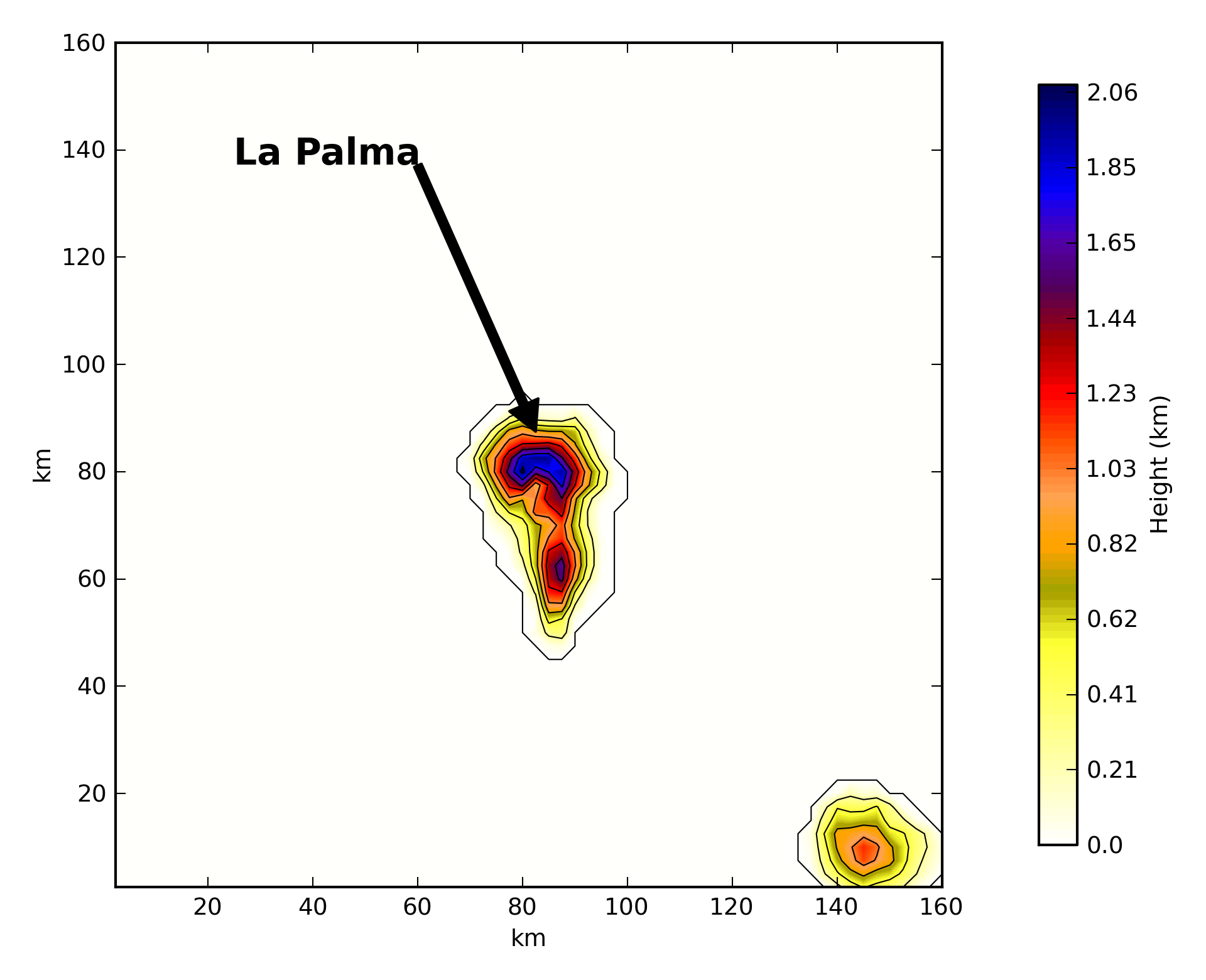}
\end{tabular}
\begin{tabular}{cc} 
\includegraphics[height=5cm]{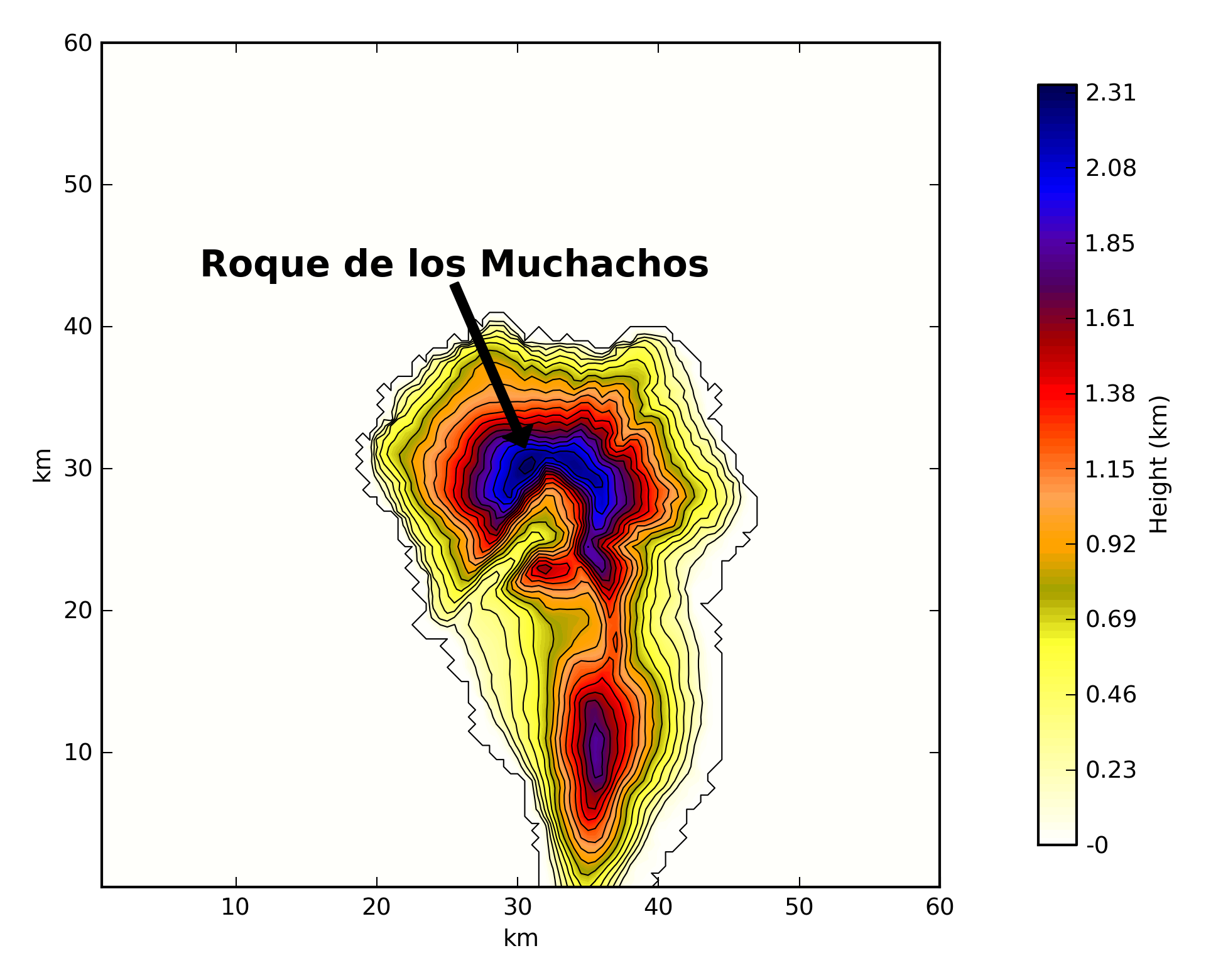} & \includegraphics[height=5cm]{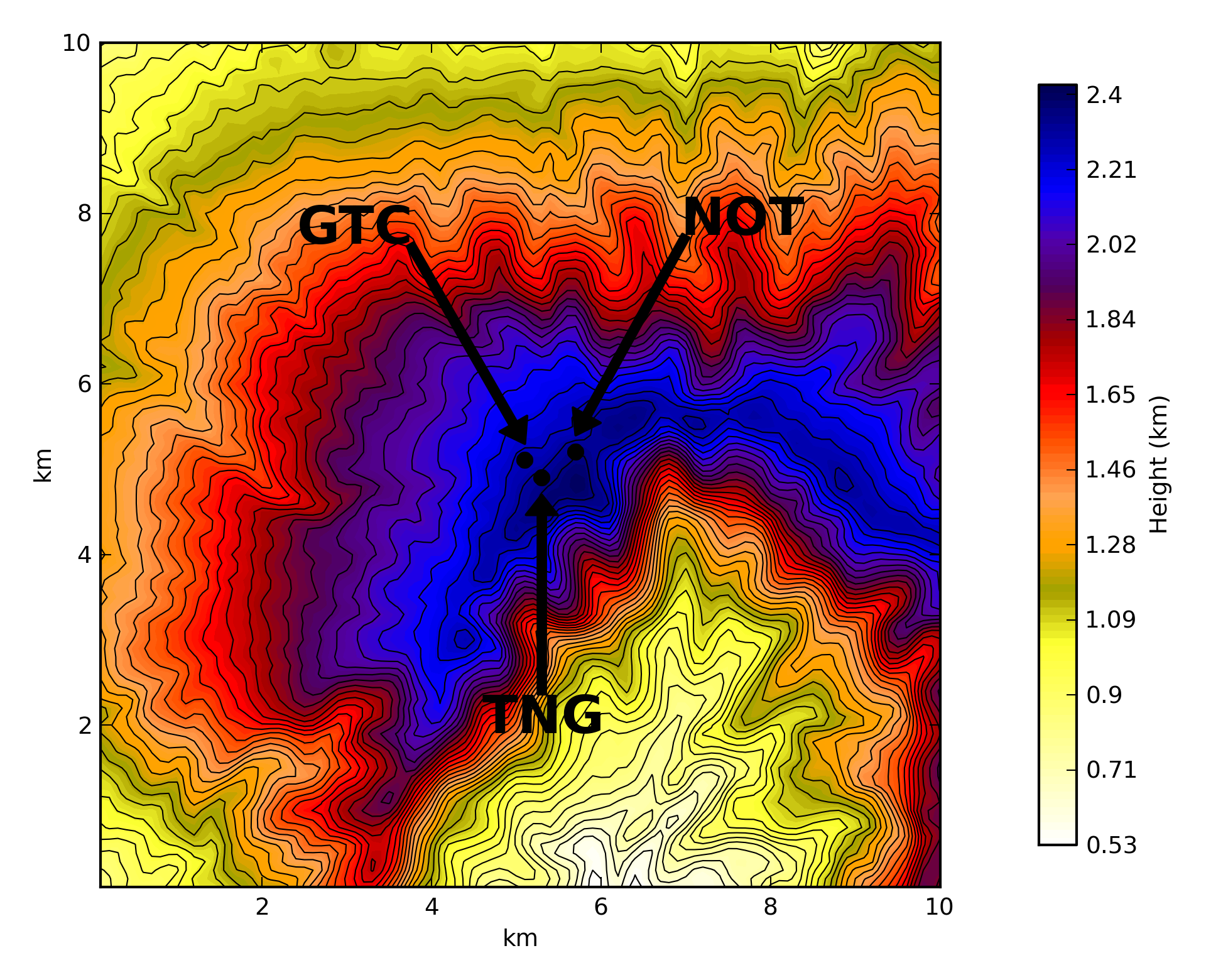}
\end{tabular}
\end{center}
\caption[example]{\label{fig:Rmaps}
From top to bottom, left to right: the 4 imbricated domains with increasing horizontal resolution (respectively $\Delta x = 10km, \Delta x = 2.5km, \Delta x = 500m, \Delta x = 100m $). Also in 4 domain (right bottom sub-figure) we indicate the location of GTC, TNG and NOT.}
\end{figure}

\begin{table}[h]
\caption{Extension of the surface of each imbricated domain of the Meso-Nh model with associated horizontal resolution.} 
\label{tab:resol}
\begin{center}       
\begin{tabular}{cccc} 
\hline
\rule[-1ex]{0pt}{3.5ex}  Domain & $\Delta$X (km) & Grid points & Domain size (km) \\
\hline
\rule[-1ex]{0pt}{3.5ex}  Domain 1 & 10 & 80x80 & 800x800 \\
\rule[-1ex]{0pt}{3.5ex}  Domain 2 & 2.5 & 64x64 & 160x160 \\
\rule[-1ex]{0pt}{3.5ex}  Domain 3 & 0.5 & 120x120 & 60x60 \\
\rule[-1ex]{0pt}{3.5ex}  Domain 4 & 0.1 & 100x100 & 10x10 \\
\hline
\end{tabular}
\end{center}
\end{table}
We used GTOPO\footnote{\url{https://lta.cr.usgs.gov/GTOPO30}} as a Digital Elevation Model for domains 1 and 2, with an intrinsic resolution of 1~km. In domains 3 and 4 we use the SRTM90\footnote{\url{http://www.cgiar-csi.org/data/srtm-90m-digital-elevation-database-v4-1}} [\cite{srtm}], with an intrinsic resolution of approximately 90~m (3 arcsec).\\
We use a 2-way interaction between the interfaces of the nested domains, so that the atmospheric flow of the innermost domain is always in a constant thermodynamic equilibrium with the outermost one, allowing for the propagation of gravity waves through the whole area mapped by the simulation.\\
The vertical grid is divided into 62 physical levels, with the first grid point placed 5~m above ground level (a.g.l.). The grid space grows with a logarithmic stretching of 20\% up to 3.5~km a.g.l., then after this threshold, the model uses an almost constant vertical grid size of $\sim$ 600~m up to 23,57 km, which is the top level of the simulation. The grid mesh deforms uniformly to adapt to the orography, so the actual size of the vertical levels can stretch in order to accommodate for the different ground level at each horizontal grid point. In Fig.\ref{fig:wethstat} we indicated the heights of each sensor placed on the different telescope and the corresponding model level that will be used as a comparison term.\\
Measurements of temperature, RH and wind speed (WS) simulated at the heights indicated in Fig.\ref{fig:wethstat} are provided with a temporal frequency equal to 1 sec. While we configured the model to reach 100m horizontal resolution in the 4th domain, for this preliminary analysis we limited ourserlves to analyse the outputs coming from the 500m horizontal resolution third domain (see Fig.\ref{fig:Rmaps}). In previous studies [\cite{lascaux2013}] we already shown than an horizontal resolution of 100~m allows a better accuracy in reconstructing the strong wind speed close to the ground. We simply note here that, because of the important steepness of the Caldera on the summit of the Roque de los Muchachos Observatory, simulations with $ \Delta x = 100m$ are hardly performable. The numerical convergence is frequently missed and we are forced to use a more relaxed horizontal resolution to respect the CLF (Courant-Friedrichs-Lewy) conditions for the stability of numerical scheme. This means that it might be a problem to run 100~m resolution simulations above La Palma. Due to the long computational times needed for such simulations, we decided to postpone that specific analysis to future works.\\

\section{MODEL VALIDATION}
\label{sec:mod_valid}
In this section we present a selection of results obtained by means of comparison between measures and model outputs for the atmospheric parameters, temperature, wind speed and relative humidity. The results are shown in term of scatter plots with standard statistical operators, defined as:
\begin{equation}
BIAS = \sum\limits_{i = 1}^N {\frac{{(Y_i  - X_i )^{} }}
{N}} 
\label{eq1}
\end{equation}
\begin{equation}RMSE = \sqrt {\sum\limits_{i = 1}^N {\frac{{(Y_i  - X_i )^2 }}
{N}} } 
\label{eq2}
\end{equation}
where $X_{i}$ are the individual observations and $Y_{i}$ the individual simulations calculated at the same time index $i$, with $1\leq i\leq N$, $N$ being the total sample size.\\
From the two latter quantities we deduce the bias-corected RMSE ($\sigma$):
\begin{equation}\sigma = \sqrt {RMSE^2 - BIAS^2}
\label{eqr}
\end{equation}
which represents the total statistical error of the model once the bias is removed.\\
We collected measurements related to two samples of 20 nights in 2011 and 2017 to be compared with model outputs so to provide results in statistical terms for the different atmospheric parameters (temperature, relative humidity and wind speed).

\begin{figure} [ht]
\begin{center}
\begin{tabular}{cc} 
\includegraphics[height=5cm]{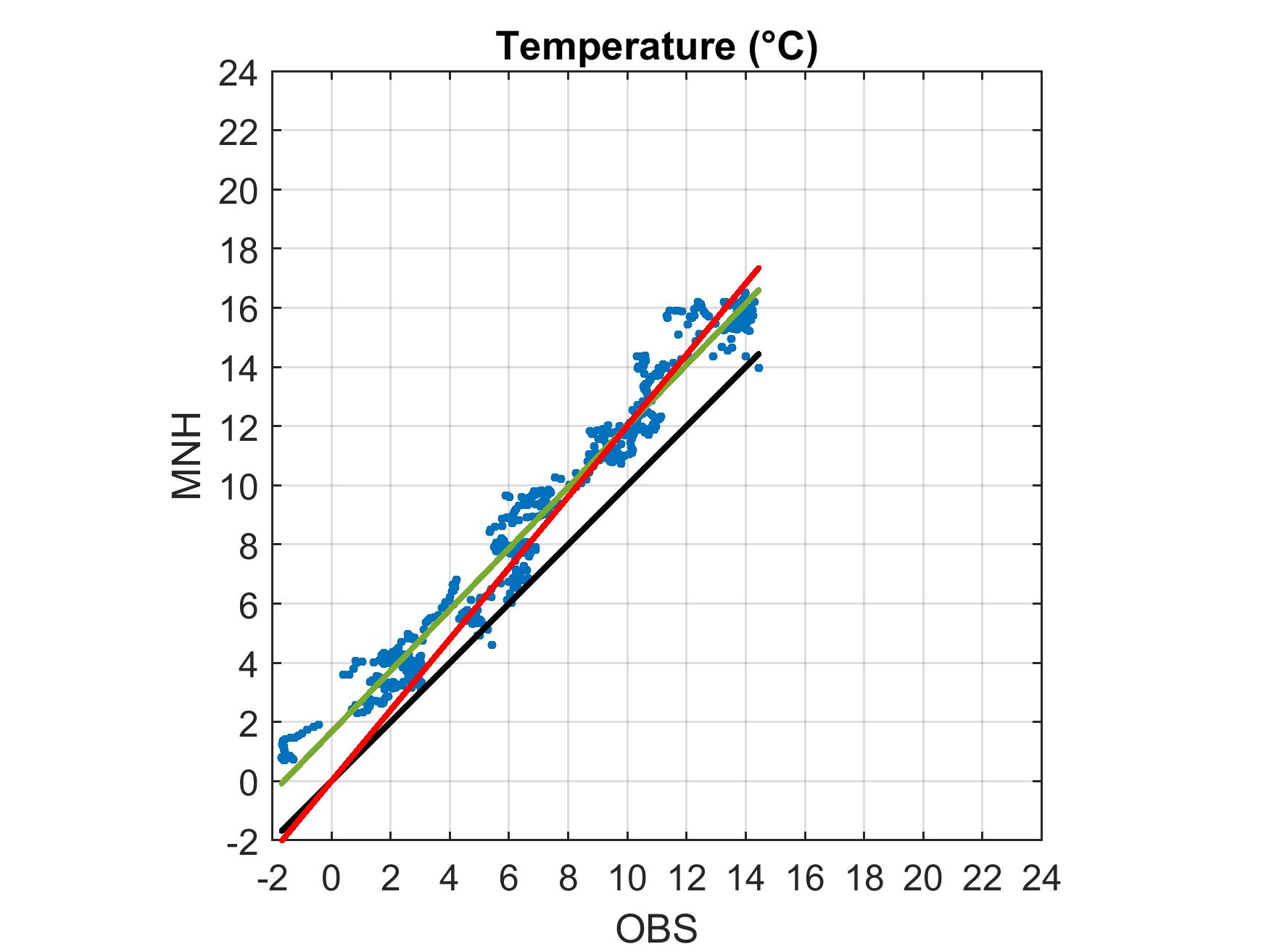} & \includegraphics[height=5cm]{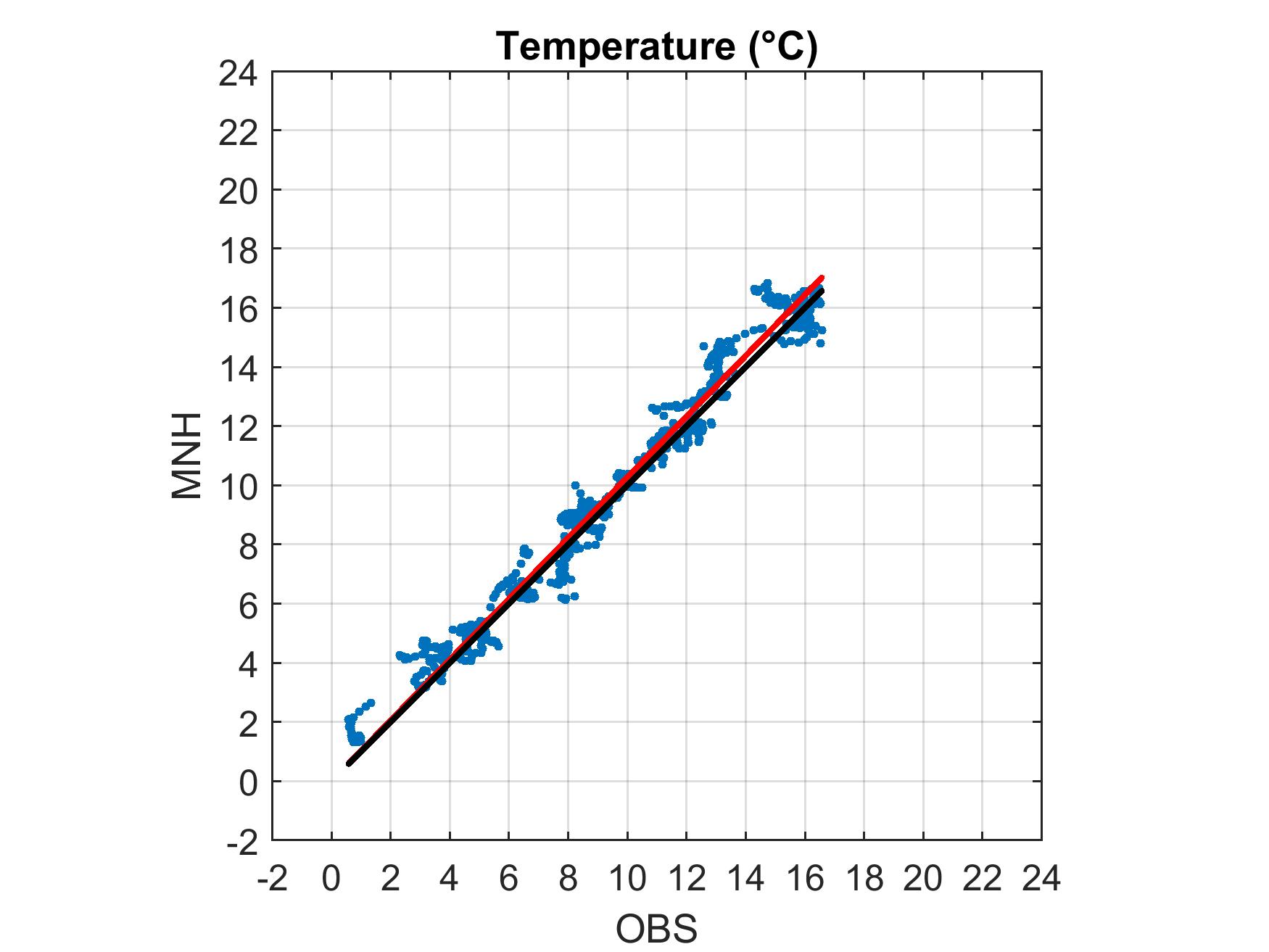}
\end{tabular}
\end{center}
\caption[example]{\label{fig:scattemp2011n}
NIGHT TIME - Left: Temperature for NOT with the 2011 sample of 20 nights. The green curve is the fit with intercept to highlight an off-set of 1.66 degrees (BIAS=1.91$^{\circ}$, RMSE=2.09$^{\circ}$, $\sigma$=0.84$^{\circ}$). Right: Temperature for TNG and the 2011 sample of 20 nights (BIAS=1.91$^{\circ}$, RMSE=2.09$^{\circ}$, $\sigma$=0.84$^{\circ}$).}
\end{figure} 

\begin{figure} [ht]
\begin{center}
\begin{tabular}{cc} 
\includegraphics[height=5cm]{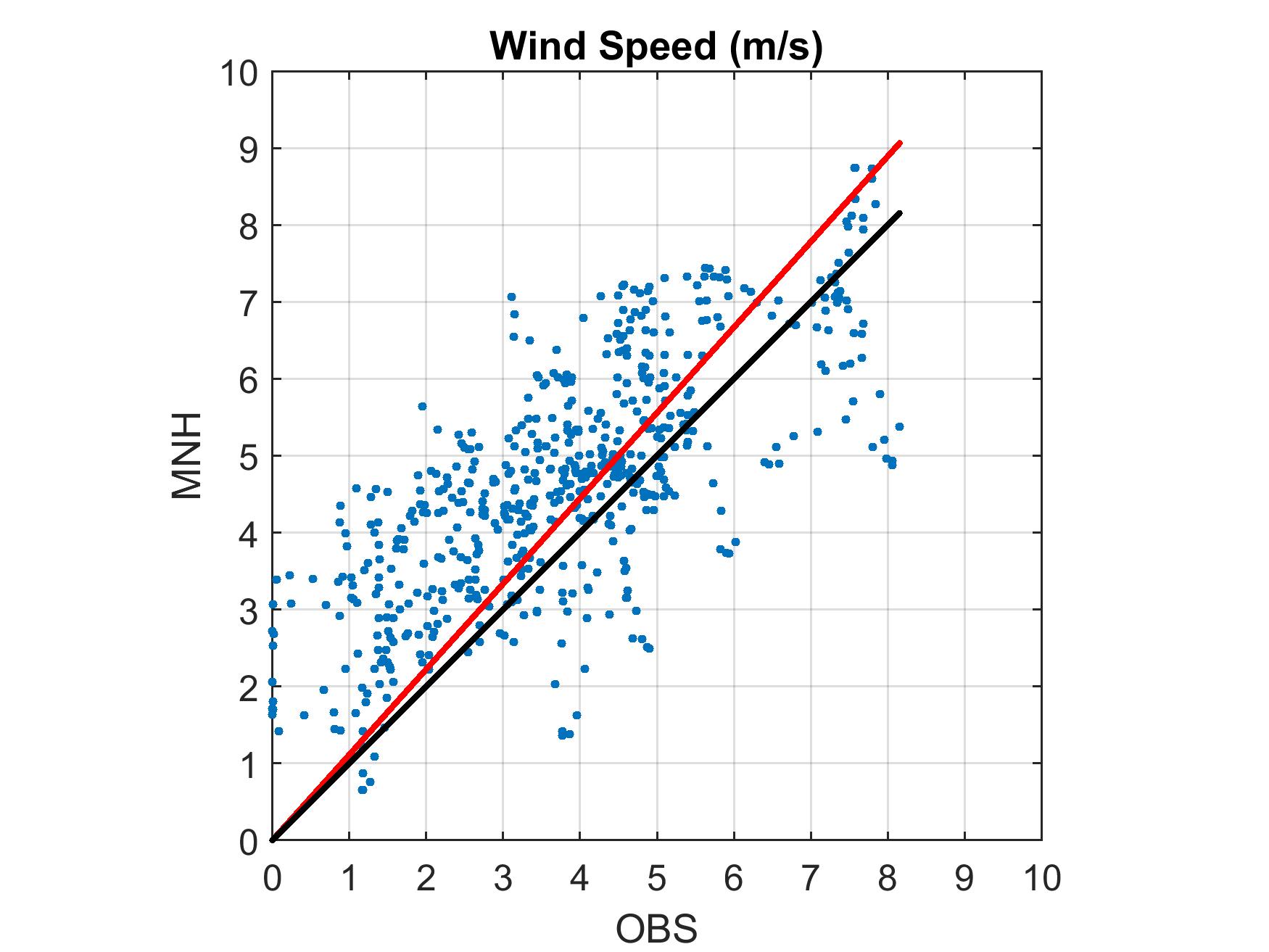} & \includegraphics[height=5cm]{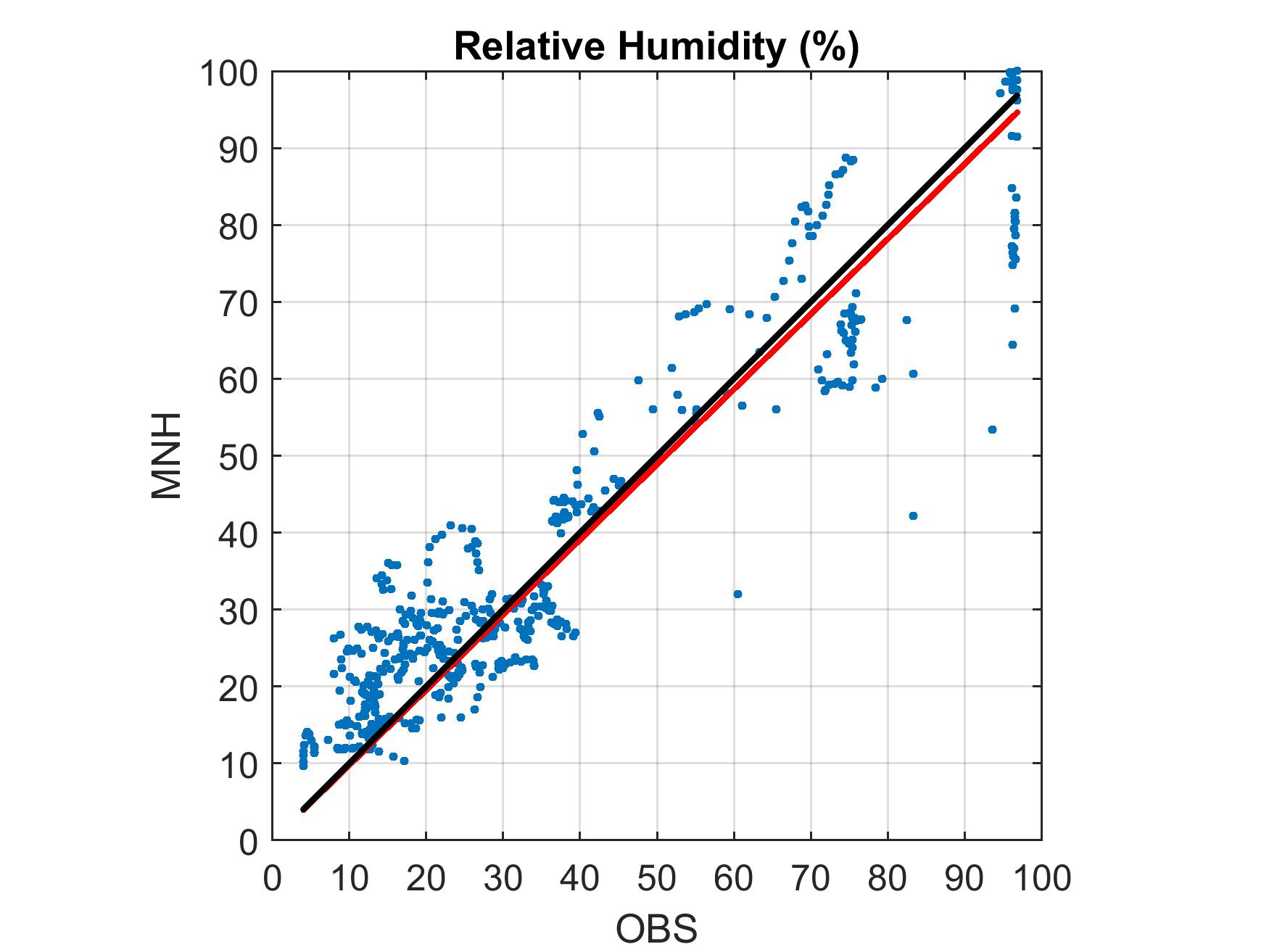}
\end{tabular}
\end{center}
\caption[example]{\label{fig:scatRHWS2017n}
NIGHT TIME - Left: Relative Humidity for GTC with the 2017 sample of 20 nights (BIAS=1.77 $\%$, RMSE=8.96 $\%$, $\sigma$=8.79 $\%$).
Right:  Wind Speed for GTC and the 2017 sample of 20 nights (BIAS=0.85 ms$^{-1}$, RMSE=1.52 ms$^{-1}$, $\sigma$=1.26 ms$^{-1}$).}
\end{figure}

In night time, for the 2011 case, we report in Fig.\ref{fig:scattemp2011n} respectively the scatter plot of observations-model outputs of the temperature for NOT (with sensor at 3~m a.g.l.) and TNG (with sensor at 5~m a.g.l.). While the TNG case provide excellent correlation we note that for the NOT case there is a clear off-set between model and measurements. Thanks to the fact that we have measurements from two different sensors related to different weather stations (TNG and NOT) we can conclude that the problem is in measurements and not in simulations. In other words, thanks to simultaneous multi-measurements, we can conclude that there is a problem on measurements of NOT (around 1.66 degrees off-set on temperature). Values of the statistical operators are BIAS=0.34$^{\circ}$, RMSE=0.80$^{\circ}$, $\sigma$=0.72$^{\circ}$ for the TNG and, BIAS=1.91$^{\circ}$, RMSE=2.09$^{\circ}$ and $\sigma$=0.84$^{\circ}$ for the NOT. The values of $\sigma$ in the two cases is comparable (0.71 and 0.84) and below 1$^{\circ}$ as we observed in other studies (\cite{lascaux2015, turchi2017}). This tells us that, assuming to correct the bias on measurements of NOT, the statistical dispersion is comparable in the two sites. For the 2017 case, always in night time, we show in Fig.\ref{fig:scatRHWS2017n} the scattering plot of the relative humidity RH (sensor at 2~m a.g.l.) and wind speed (sensor at 10~m a.g.l.) for GTC. Values of the statistical operators are BIAS=1.77 $\%$, RMSE=8.96 $\%$ and $\sigma$=8.79 $\%$ for the RH and  BIAS=0.85 ms$^{-1}$, RMSE=1.52 ms$^{-1}$ and $\sigma$=1.26 ms$^{-1}$ for the wind speed. We note a general coherence with previous studies regarding Cerro Paranal and Mount Graham [\cite{lascaux2015, turchi2017}]. In some cases results seems even better but the sample is still at present small. It is important to increase the number of nights to quantify as the statistical operators change. The wind speed seems to be correctly estimated and we do not observe, at least with this preliminary sample of nights, underestimates of the model as it was the case above Cerro Paranal.

\begin{figure} [ht]
\begin{center}
\begin{tabular}{cc} 
\includegraphics[height=5cm]{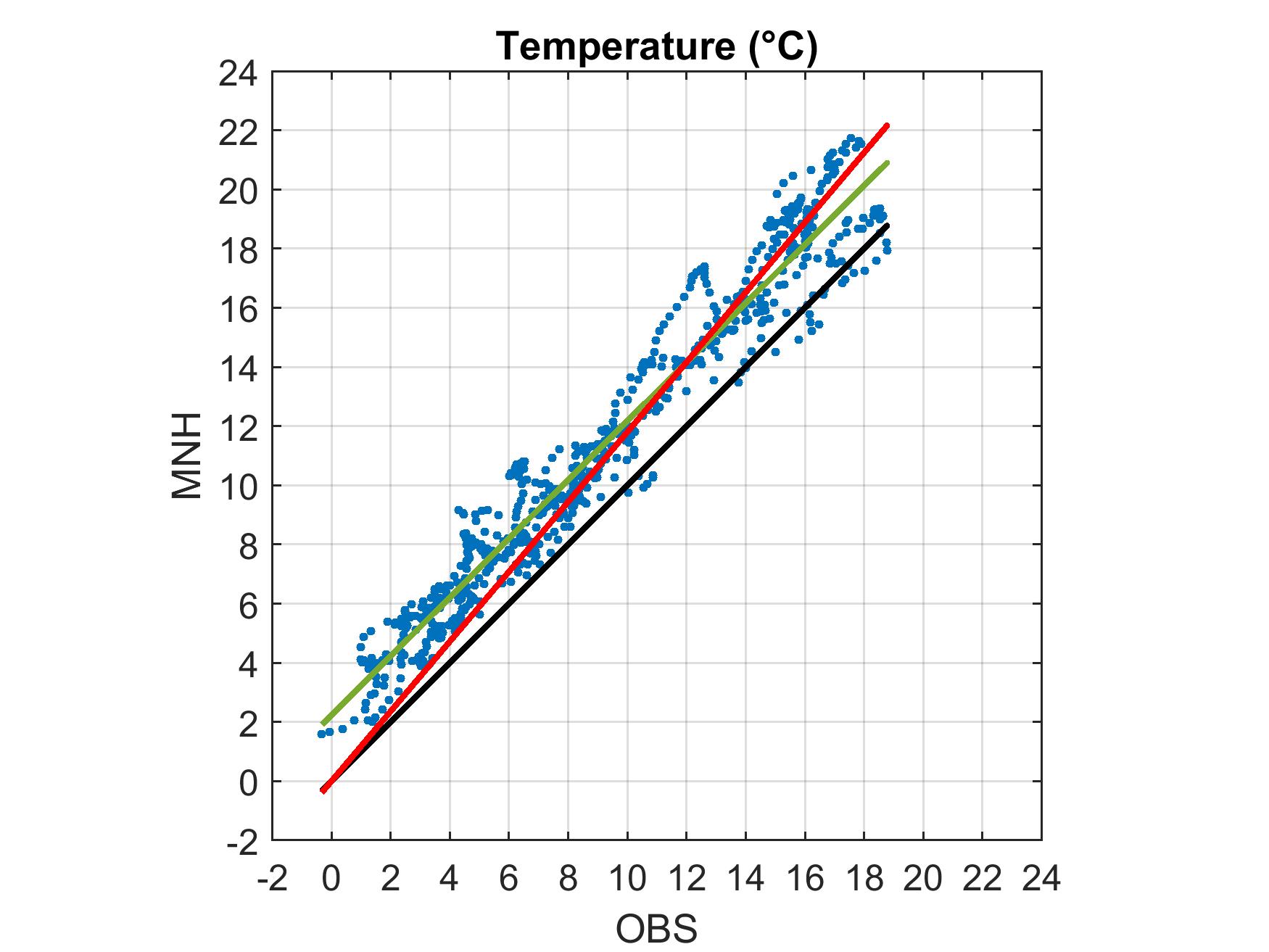} & \includegraphics[height=5cm]{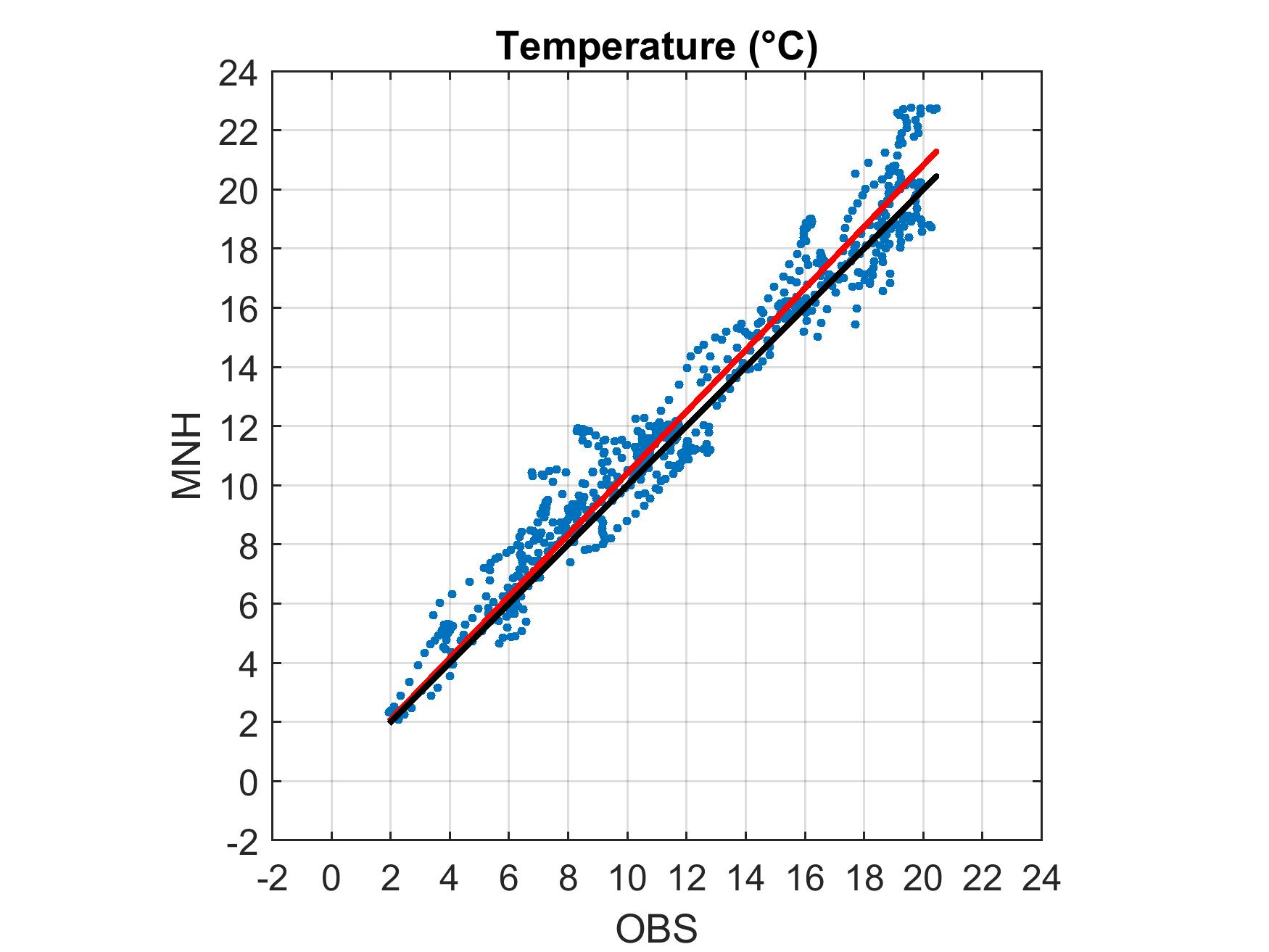}
\end{tabular}
\end{center}
\caption[example]{\label{fig:scattemp2011d}
DAY TIME - Left: Temperature for NOT with the 2011 sample of 20 nights. The green curve is the fit showing an off-set of 2.24 degrees (BIAS=2.18$^{\circ}$, RMSE=2.47$^{\circ}$ and $\sigma$=1.17$^{\circ}$). Right:  Temperature for TNG and the 2011 sample of 20 nights (BIAS=0.61$^{\circ}$, RMSE=1.27$^{\circ}$, $\sigma$=1.12$^{\circ}$).}
\end{figure}

\begin{figure} [ht]
\begin{center}
\begin{tabular}{cc} 
\includegraphics[height=5cm]{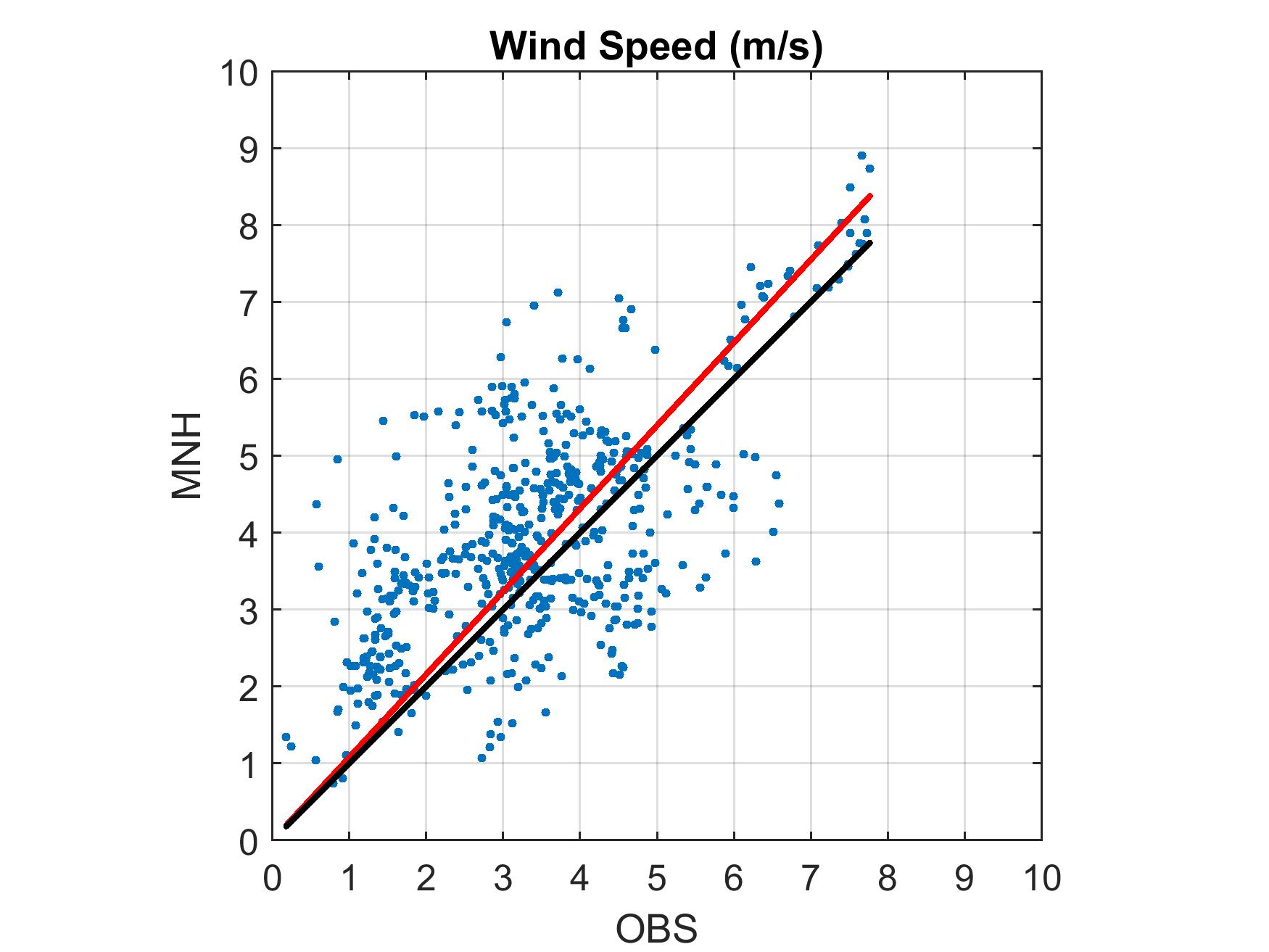} & \includegraphics[height=5cm]{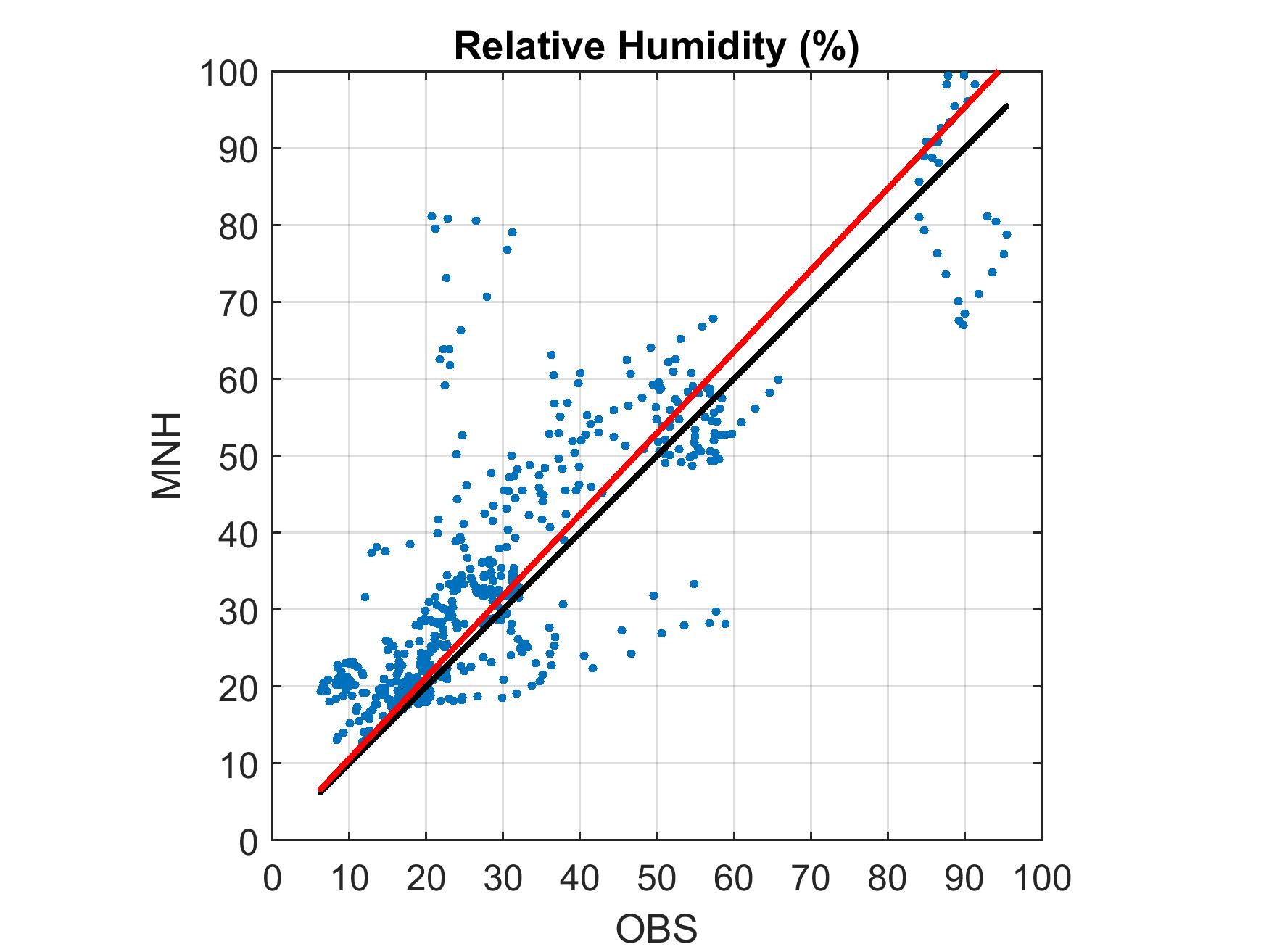}
\end{tabular}
\end{center}
\caption[example]{\label{fig:scatRHWS2017d}
DAY TIME - Left: Relative Humidity for GTC with the 2017 sample of 20 nights (BIAS=4.85 $\%$, RMSE=11.98 $\%$ and $\sigma$=10.95 $\%$).
Right:  Wind Speed for GTC with the 2017 sample of nights (BIAS= 0.56ms$^{-1}$, RMSE=1.35 ms$^{-1}$, $\sigma$=1.23 ms$^{-1}$).}
\end{figure}

In day time, for the 2011 case, we report in Fig.\ref{fig:scattemp2011d} respectively the scatter plots of observations vs. model outputs of the temperature for NOT and TNG. We observe also in day time an off-set on measurements of NOT. The off-set is slightly larger in day-time (around 2.24 degrees) but this might be a statistical effect considering the sample is still relatively small. A richer statistic is necessary to better estimate the off-set on night and day time. Values of the statistical operators are BIAS=0.61$^{\circ}$, RMSE=1.27$^{\circ}$ and $\sigma$=1.12$^{\circ}$ for the TNG and BIAS=2.18$^{\circ}$, RMSE=2.47$^{\circ}$ and $\sigma$=1.17$^{\circ}$ for the NOT. Values of $\sigma$ seems a little bit larger in day time than in night time. However a larger sample is necessary to be able to confirm this trend. On this sample of 20 nights we observe an average value of excess $\Delta\sigma$ $\sim$ 0.3$^\circ$ C in the central segment of day [12:00-14:00] LT with respect to the tails [08:00-09:00] LT plus [17:00-18:00] LT. This seems to indicate that the larger $\sigma$ during the day time might be caused by the solar radiation.
For the 2017 case, in day time, we show in Fig.\ref{fig:scatRHWS2017d} the scattering plots respectively of the relative humidity RH and wind speed for GTC. Values of the statistical operators are BIAS=4.85 $\%$, RMSE=11.98 $\%$ and $\sigma$=10.95 $\%$ for the RH and BIAS= ms$^{-1}$, RMSE= ms$^{-1}$ and $\sigma$= ms$^{-1}$ for the wind speed. For what concerns RH, $\sigma$ is slightly larger in day time than in night time while the wind speed is comparable to what has been observed on night time. A larger sample of nights is suitable to confirm these preliminary results. \\

\section{CONCLUSIONS}
\label{sec:concl} 
In this paper we present preliminary results of the on-going 24h validation study (night and day) on Roque de Los Muchachos observatory site. Results shown in this contribution refer to two samples of 20 nights, one on 2011 and the other on 2017. We compare outputs of the mesoscale atmospherical model Meso-Nh related to atmospheric paramaters (temperature, wind speed and relative humidity) with observed values taken near the ground level. Measurements have bene obtained with weather station instruments located near the telescopes that we take as a reference. Our preliminary results indicate (1) a general coherent model performances with previous studies in the night time regime in other astronomical sites such as Mt.Graham and Cerro Paranal [\cite{lascaux2015, turchi2017}] even if we envisage to increase the statistical sample of nights to provide a more accurate estimate; (2) model performances during the day time are similar to the night time regime. A slightly larger $\sigma$ is observed for the temperature and RH.  Results to be confirmed with a richer sample of observations. The next step is the analysis of optical turbulence in night and day regime. The latter is motivated by astronomical applications (solar telescopes) and also satellite communications.

\acknowledgments  
We acknowledge IAC (Julio Castro Almazan and Luzma Montoya) and TNG (Albar Garcia de Guturbai) for providing us in situ measurements of the atmospheric parameters.


\end{document}